\documentclass[]{article}
\usepackage[top=1cm, left=2cm, bottom=2cm, right=1cm]{geometry}
\usepackage[english]{babel}
\usepackage{amssymb}
\usepackage{amsmath}
\usepackage{graphicx}
\usepackage{color}
\usepackage{mathtools}
\usepackage{authblk}

\usepackage{tikz}
\usetikzlibrary{positioning,arrows}
\usetikzlibrary{decorations.pathmorphing}
\usetikzlibrary{decorations.markings}

\tikzset{
vector/.style={thick,double,draw=black, postaction={decorate},
    decoration={markings,mark=at position .6 with {\arrow[black]{triangle 45}}}},
axial/.style={thick,double,densely dashed,draw=black, postaction={decorate},
    decoration={markings,mark=at position .6 with {\arrow[black]{triangle 45}}}},
gluon/.style={decorate, draw=black,
    decoration={coil,aspect=0.3,segment length=5pt,amplitude=3pt}},
pseudo/.style={thick, dashed, draw=black, postaction={decorate},
    decoration={markings,mark=at position .6 with {\arrow[black]{triangle 45}}}},
scalar/.style={thick,draw=black, postaction={decorate},
    decoration={markings,mark=at position .6 with {\arrow[black]{triangle 45}}}},
cut/.style={very thick, densely dashed, draw=gray}
 }

\title{A triangle singularity and the LHCb pentaquarks}
\author{M.~\textsc{Mikhasenko}
\thanks{mikhail.mikhasenko@hiskp.uni-bonn.de}
}
\affil{Universit\"at Bonn, Helmholtz-Institut f\"ur Strahlen-
  und Kernphysik, 53115 Bonn, Germany}
\date{July 22, 2015 }

\begin{document}
\maketitle

\begin{abstract}
In this article, a possible interpretation of the pentaquark candidates recently observed at LHCb is given. We show that the reaction dynamics and the peculare kinematics situation can produce the peak in the spectrum and the sharp phase motion. The mechanism called triangle singularity likely causes the appearance of the new pentaquarks candidates.
\end{abstract}

\section{Introduction}

The search for pentaquarks has a long-time history. The light states have been predicted and discovered, but then was closed several years later \cite{Hicks:2012zz}. Recently, the resonace-like signals observed by the LHCb collaboration brought back the attention to the subject. The decay of $\Lambda_{b}$ was studied in the mode $K^- \, J/\Psi\,p$ \cite{Aaij:2015tga}. Even the raw $J/\Psi\,p$ spectrum shows the peak around $4450\,$MeV. 
Impressive work on splitting the observed intensity of the process to different components employing the isobar model has been done to show that known resonances are not enough to describe the mass- and angular-distributions. Thus, the two additional resonances $P_c(4380)$ and $P_c(4450)$ have been introduced. The authors carefully studied the complex amplitude and showed that it behaved as a circle in the Argon-diagram. 
The quark content of the pentaquark could be $uudc\bar{c}$. That states might open new region in the particle physics. The such pentaquark $uudc\bar{c}$ could mix with the proton explaining the intrinsic charm effect \cite{Mikhasenko:2012km}. 
The new observed resonace-like signals were also immediately interpreted as a meson-baryon molecules be several groups \cite{Roca:2015dva,Mironov:2015ica}.

We are giving another interpretation, in which the enhancement in the spectrum with the phase motion does not correspond to a resonance and a pole in the amplitude. Rather, it is caused by the rescattering of known particles in the triangle diagram in such way to introduce a logarithmic branching point in the amplitude very close to the physical region. 
The process with rescattering of kaons in a triangle diagram was employed to describe $C$-meson ($\phi\pi$ final state) in ninetieth \cite{Achasov:1989ma} for Lepton-F experiment \cite{Bityukov:1986yd}, $a_1(1420)$ in ref.~\cite{Ketzer:2015tqa} ($3\pi$ final state), observed by the COMPASS and VES experiments \cite{Adolph:2015pws,Khokhlov:2014nha}, the structure of $\eta(1405)/\eta(1440)$ in ref.~\cite{Wu:2011yx} for BESIII experiment.

In this paper, we are presenting a study of the reaction $\Lambda_b \to D_{sJ}^{*-} \Sigma_{c}^+ \to K^- D^{*0} \Sigma_{c}^+ \to K^-\, J/\Psi \, p$, showing the importance of the triangle rescattering in unredstanding of the $J/\Psi \, p$ stectrum.
The similar idea is discussed in the ref.~\cite{Liu:2015fea,Guo:2015umn}, considering different loops.

The paper will be structured in this way that we will first demonstrate that the reaction $\Lambda_b \to D_{sJ}^{*-} \Sigma_{c}^+ \to K^- D^{*0} \Sigma_{c}^+ \to K^-\, J/\Psi \, p$ is well fitted by the quark model. The second section is dedicated to the calculation. There we will explain in details why the reaction were chosen for the consideration. In the third section we will show the results of our calculation and compare them to the resonances parametrization obtained by LHCb. At the end, the issues of the approach will be discussed.

\section{Quark level}

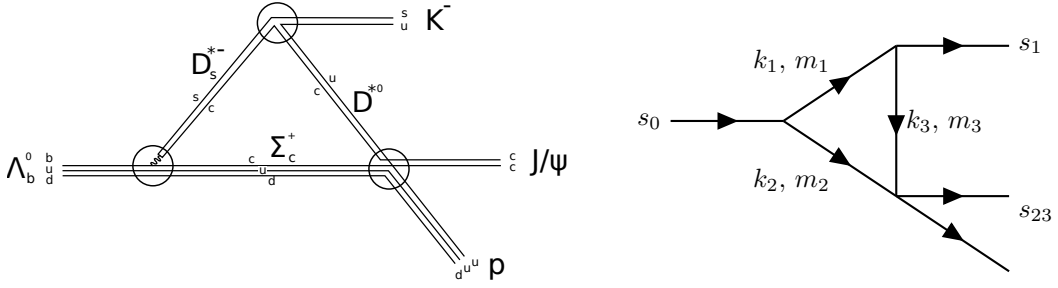
\begin{figure}[h]
\centering
\includegraphics[width=0.4\textwidth]{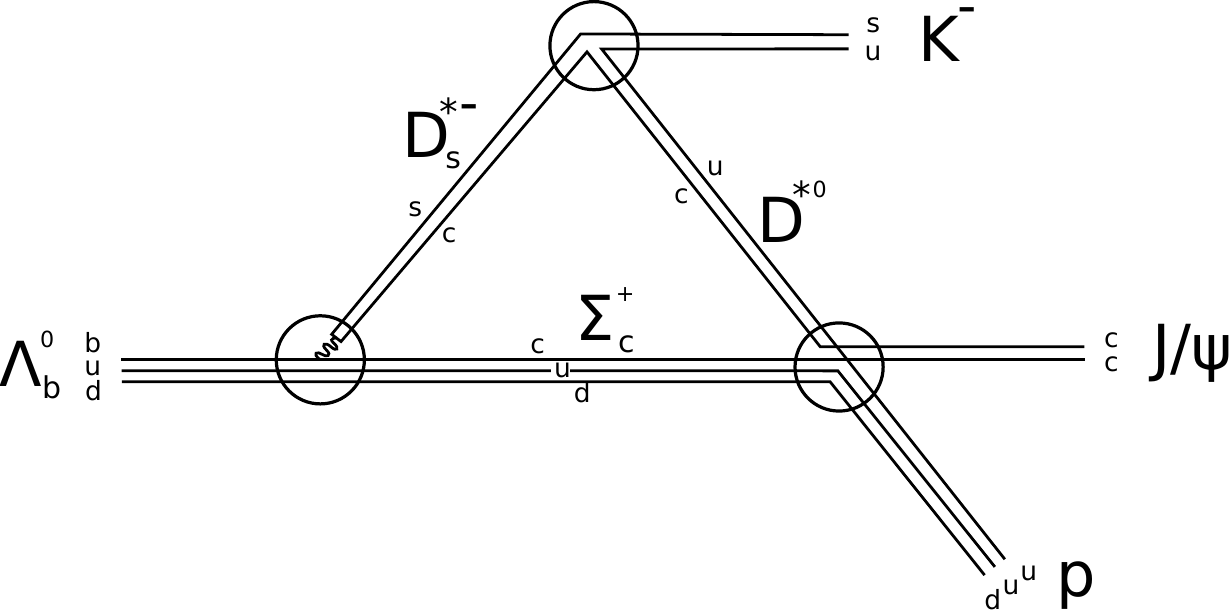} \qquad
  \begin{tikzpicture}[node distance=1cm and 1.5cm, baseline=-60.]
    \coordinate[label=left:{$s_0$}] (a1); 
    \coordinate[right=of a1] (a2); 
    \coordinate[above right=of a2] (b1); 
    \coordinate[below right=of a2] (b2); 
    \coordinate[right=of b1,label=right:{$s_1$}](c1); 
    \coordinate[right=of b2,label=below right:$s_{23}$] (c2);
    \coordinate[below right=of b2] (d2);
 
    \draw[scalar] (a1) -- (a2); 
    \draw[scalar] (a2) -- node[label=above left:{$k_1$, $m_1$} ] {} (b1); 
    \draw[scalar] (a2) -- node[label=below left:{$k_2$, $m_2$} ] {} (b2); 
    \draw[scalar] (b1) -- node[label=right:{$k_3$, $m_3$}] {} (b2); 
    \draw[scalar] (b1) -- (c1); 
    \draw[scalar] (b2) -- (c2);
    \draw[scalar] (b2) -- (d2);
  \end{tikzpicture}
\caption{Left: quark diagram of considered process; right: used notations for the calculation.}
\label{fig:quark}
\end{figure}

Based on the quark composition of the final state particles we may conclude that the weak process takes place with a $b \to c$ quark transition and an appearance of a strange quark paired with a $c$-quark. The direct diagrams of such processes are shown in the ref.~\cite{Aaij:2015tga}. The state $D_s^- \Sigma_c^+$ has the same quark content and the decay $\Lambda_b^0\to D_s^- \Sigma_c^+$ is also allowed by quantum numbers conservation. The $D_s^{\star-}$ is only decaying to $K^- D^{\star0}$ via the strong interaction. Then the reaction $D^{\star0} \Sigma_c^+ \to J/\Psi \, p$ can happen. The quark-cartoon of the process is shown in the Fig.~\ref{fig:quark}.

\section{The location of the branching point}
First, consider the process of rescattering of scalar particles with infinite small widths, which is shown at the right side of Fig.~\ref{fig:quark}.
The amplitude for the tringle diagram can be written as:
\begin{equation}
A_\Delta^{\text{sc}}(s_0,s_1,s_{23}) = \int \frac{\mathrm{d}^4 k_{1,\mu}}{(2\pi)^4 i} \frac{1}{\Delta_1 \Delta_2 \Delta_3},
\end{equation}
where $\Delta_i = m_i^2-k_{i,\mu}^2-i\varepsilon$ corresponds to the inverse scalar propagator of the particle $i$ and $k_{1,\mu}$ is the momentum in the loop. $s_0,s_1,s_{23}$ are external invariants denoted in Fig.~\ref{fig:quark}.
The integral results in a singularity when the integration domain is pinched by singularities of the integrand. The mathematics was firstly developed by Landau \cite{Landau:1959fi} and is well described in the book \cite{Eden:2002en}. 
We are using the Feynman approach to calculate the integral. The expression for $A_{\Delta}^{\text{sc}}$ is giving in the ref.~\cite{Ketzer:2015tqa}.
Sometimes, the dispersion representation of the integral is used for the calculation of it and the position of singularity \cite{Aitchison:1964zz,Lucha:2006vc,Szczepaniak:2015eza}.

We are discussing the amplitude as a function of $s_{23}$, assuming that the other variables are fixed. The positions of singularities  in the compex $s_{23}$ plane are given by Eq.~\ref{eq:system}.
\begin{equation} \label{eq:system}
\left\{
\begin{array}{l}
k_{i,\mu}^2 = m_i^2, \quad i = 1\dots 3\\
x k_{1,\mu} - y k_{2,\mu} + z k_{3,\mu} = 0.
\end{array}
\right.
\end{equation}
Here, $k_{i\mu}$ is four-momentum of a particle in the loop, $m_i$ is the corresponding mass, $i=1,2,3$. $x,y,z$ are Feynman parameters, which is unknown in the equation. The system of equations is overdetermined. Thus, firstly, we require the system to be solvable, that determins the position of singularities.
Whether one of them is at the physical region (on the real axis) can be checked via the conditions:
\begin{equation} \label{eq:check}
\left\{
\begin{array}{l}
x+y+z = 1,\\
x,y,z > 0.
\end{array}
\right.
\end{equation}
The singularities correspond to the kinematics when all particles are on mass shell and all 4-momentum conservation laws are hold. The check conditions are valid if $D^{\star0}$ can catch up $\Sigma_c$ (catch up condition) \cite{Aitchison:1964zz}.

\begin{figure}[h]
\centering
\includegraphics[width=0.4\textwidth]{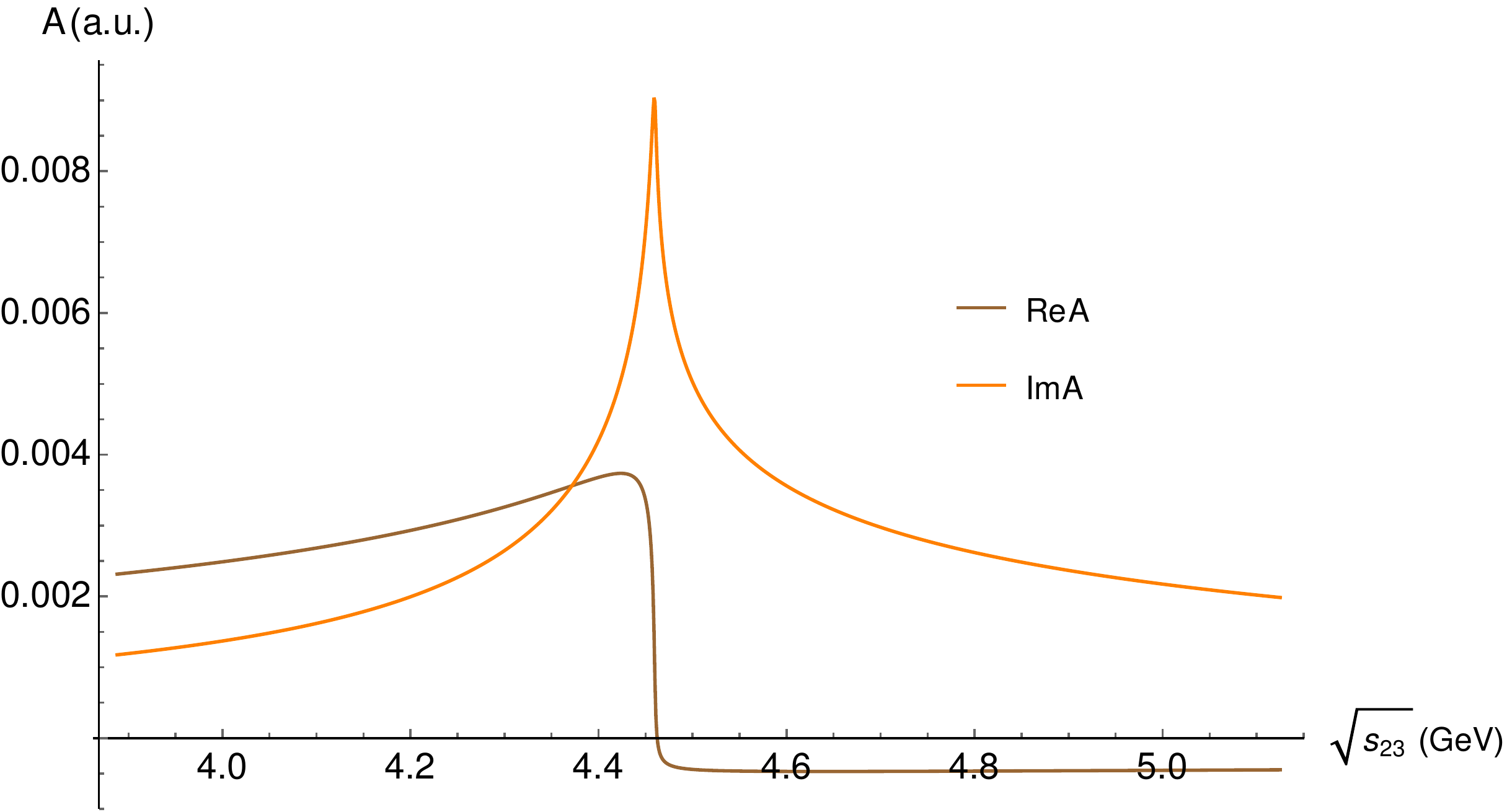}
\caption{The real and the imaginary part of the amplitude $A_\Delta^{\text{sc}} (s_{23})$ (see the text).}
\label{fig:reim}
\end{figure}

The positions of the singularities are defined by external invariants as well as the internal masses in the loop. 
In case of $\Lambda_b^0 \to K^-\, J/\Psi\, p$ there are many possible combinations of particles in the triangle. To find the leading process (the logarithmic branch point is at closest position to the physical region) we use the following logic. 
It can be shown that the position of the singularity $s_{23}^{\text{sing}}$ is close to the threshold in the corresponding channel $m_2+m_3$. 
The mass of the narrow pentaquark candidate is $4.45\,$GeV, while the mass sum of $\Sigma_c$ and $D^{\star 0}$ is $4.46\,$GeV. All different combinations of known pair meson-baryon are farther away. Thus, we use $\Sigma_c$ and $D^{\star 0}$ as a pair to rescatter to $J/\Psi \, p$.
While the masses of particles $2$ and $3$ are fixed, we can evaluate the mass of particle $3$ to have the singularity at the physical region. Analysis of Eq.~\ref{eq:system} gives the answer $m_3 \in (3.07, 3.17)\,$GeV. The lower limit comes from Eq.~\ref{eq:check}, the higher limit is just $m_{\Lambda_b^0} - m_{\Sigma_c^+}$. 
A meson with $s\bar c$ content with the mass closed to the obtained interval is $D_{sJ}^-$, based of PDG \cite{Agashe:2014kda}.

The value of masses used in the calculation is the following: 
$M_0 = m_{\Lambda_b^0} = 5.619\,$GeV, 
$M_1 = m_{K^-} = 0.494\,$GeV, 
$m_1 = m_{D_{sJ}^{\star-}} = 3.044\,$GeV, 
$m_2 = m_{\Sigma_c^+} = 2.453\,$GeV, 
$m_3 = m_{D^{\star0}} = 2.007\,$GeV.


The situation when the logarithmic singularity appears in the physical region breaks unitarity and it is unphysical. Such confusion is caused by our assumption about infinite small widths of the interacting particles. The finite widths of the recattering particle remove singularity slightly away from the real axis, making the amplitude smooth. The simplest way to evaluate such an amplitude is just replace $i \varepsilon$ on the propagators to $i m_i \Gamma_i$. The more complicated substitution with the energy dependent width would give the same result, because the widths are small. The width of $\Sigma_c^+(2455)$ has not been measured, but it seems to be the biggest width of the particle in the triangle. Since the value is unknown, We use the same as for the $\Sigma_c^0$, namely $\Gamma_{\Sigma_c^+} = 2.2\,$MeV. 
When the branching point are not in the physical region, but close to (like in our situation), the finite widths just wash out the amplitude. The real and the imaginary part of the amplitude are shown in Fig.~\ref{fig:reim}.

\section{Mass distribution}	

We compare the $J/\Psi\, p$ peaks from the experimental fit with our amplitude. For that we construct the differential width $\mathrm{d}\Gamma_{\Lambda_b \to K^-\, J/\Psi\, p}/\mathrm{d}s_{23}$ for the amplitude with the triangle rescattering $\mathrm{d}\Gamma_{\Delta}/\mathrm{d}s_{23}$ and with Breit-Wigner functions from the experimental fit.
The differential phase space of the $3$-body final state is denoted as $\mathrm{d}\Phi_3/\mathrm{d}s_{23}$ and is given by Eq.~\ref{eq:ph}.
\begin{equation} \label{eq:ph}
\frac{\mathrm{d}\Phi_3}{\mathrm{d}s_{23}} = \frac{1}{2\pi} \, \frac{1}{8\pi} \frac{\lambda^{1/2}(M_0^2,m_1^2,s_{23})}{M_0^2} \, \frac{1}{8\pi} \frac{\lambda^{1/2}(s_{23},m_2^2,m_3^2)}{s_{23}},
\end{equation}
where $\lambda(x,y,z) = x^2 + y^2 + z^2 - 2(x y + y z + z x)$ is the Kallen function. 
The differential width can be calculated as Eq.~\ref{eq:width}.
\begin{equation} \label{eq:width}
\frac{\mathrm{d}\Gamma_{\Lambda_b\to K^- J/\Psi p}}{\mathrm{d}s_{23}} =  
\frac{1}{2M_0}\left| A_{\Lambda_b\to K^- J/\Psi p} \right|^2 \frac{\mathrm{d}\Phi_3}{\mathrm{d}s_{23}},
\qquad
\frac{\mathrm{d}\Gamma_{\Delta}}{\mathrm{d}s_{23}} =  
\frac{1}{2M_0}\left| A_\Delta \right|^2 \frac{\mathrm{d}\Phi_3}{\mathrm{d}s_{23}}.
\end{equation}

\begin{figure}[h]
\centering
\includegraphics[width=0.5\textwidth]{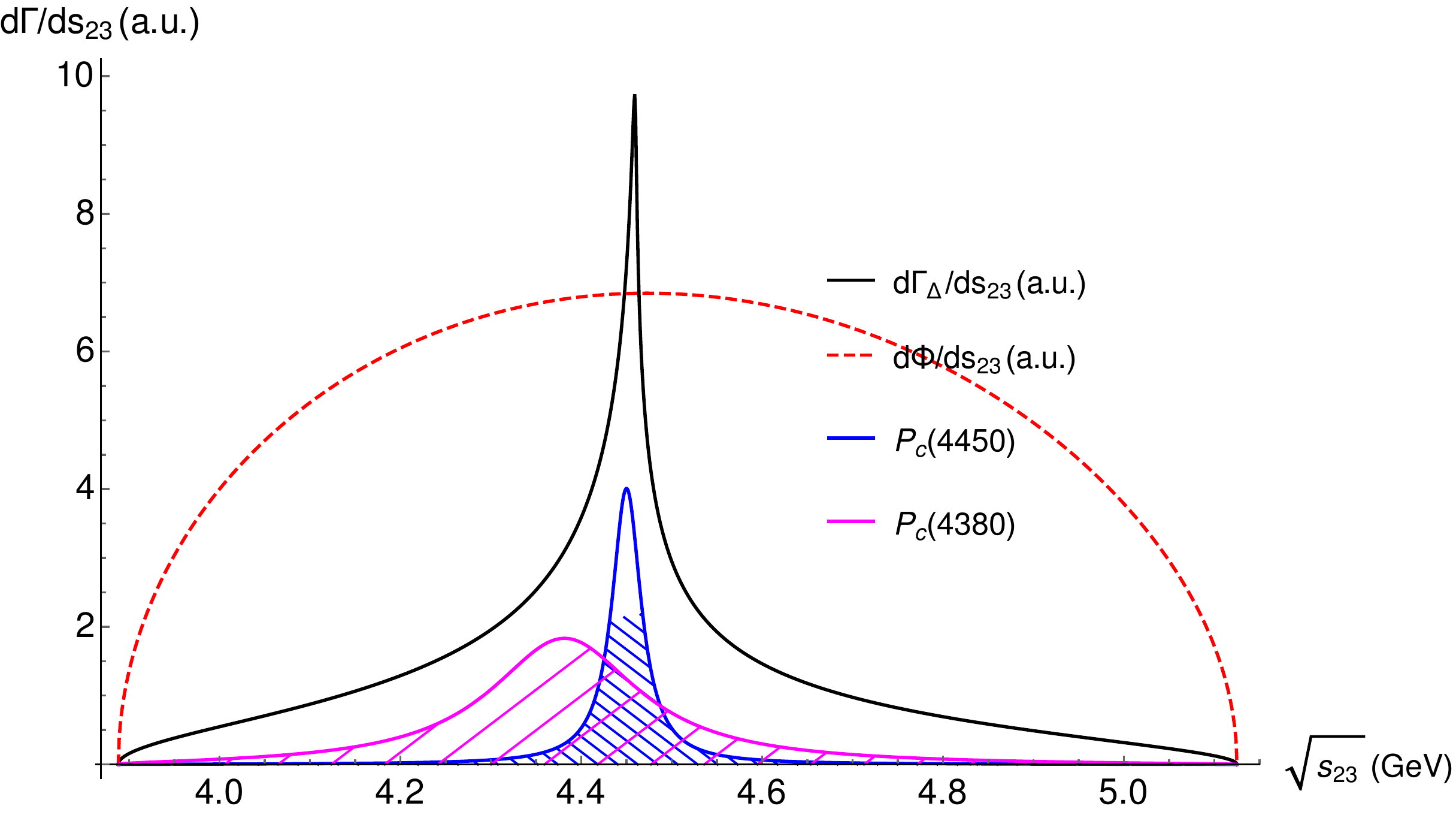}\qquad
\includegraphics[width=0.25\textwidth]{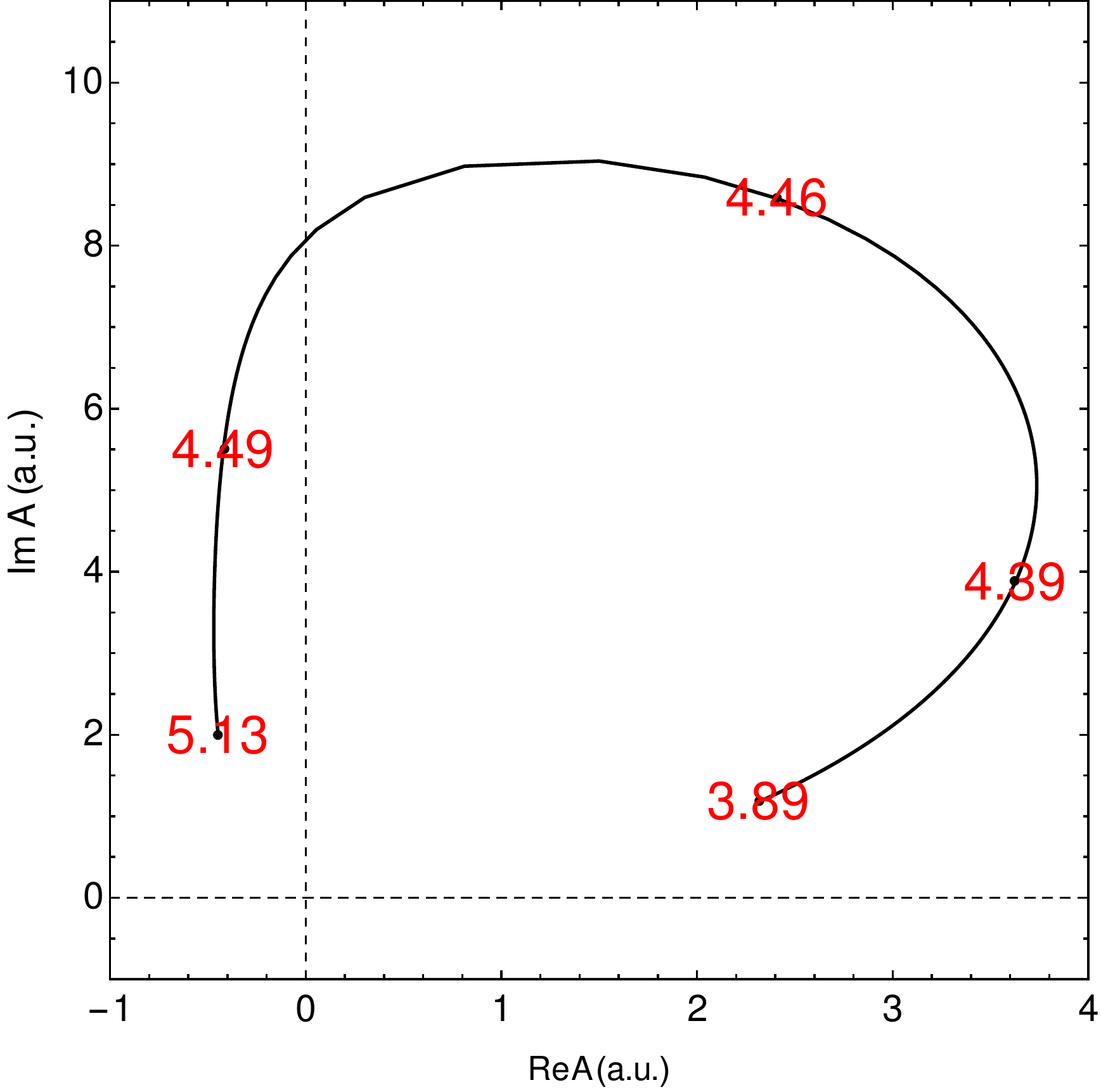}
\caption{Left plot: the differential width for the triangle diagram (black solid line), the obtained by LHCb resonance peaks (blue and magenta), the phase space are shown by red dashed line; right plot: the argon diagram (Re\,$A_\Delta^{\text{sc}}$ vs Im\,$A_\Delta^{\text{sc}}$) for the amplitude with the triangle diagram. The red numbers indicate the value of $\sqrt{s_{23}}$ at the points.}
\label{fig:gamma}
\end{figure}

The differential width for our amplitude is plotted in Fig.~\ref{fig:gamma} as a black solid line. The resonances with the parameters obtained at LHCb fit are plotted as blue and magenta lines.
The Argon diagram for our amplitude is shown in Fig.~\ref{eq:width}.

The intensity of the triangle rescattering process has a peak at the position of narrow pentaquark candidate. Left shoulder looks similar to second pantaquark candidate. The phase of amplitude $A_\Delta^{\text{sc}}$ makes a circle at the Argon diagram, similar to resonance behaviour. 

\section{Conclusion}
The narrow resonance-like structure in the spectrum $J/\Psi\,p$ can be produced by the logarithmic branching point in the amplitude. 
The discription of the effect does not require new poles for binding states.
The position of the peak and the phase motion are determined by the masses of the rescattering particles. The applied approach shows that the observed peaks and the phases likely originate from the rescattering. 

Nevertheless, our consideration is very simple. We would like to discuss a list of points, which have to be done more accurate to finally clarity the effect. 
\begin{enumerate}
\item {\it Spin, angular momentum structure, parametrization of the vertexes}: we have calculated the amplitude with the scalar particles, which is not correct. The spin of particles in the loop and non-zero the angular momentum indeed change the shape of the distribution, but not the  position of the logarithmic branching point.
The interaction $D^{\star0} \Sigma_c \to J/\Psi p$ could differ to constant interection father from threshold.
\item {\it Other processes and higher order rescattering}: the calculation of full amplitude with three body final state is rather a complicated problem. All possible two body rescattering was taking into accout for the reactions where only elastic scattering takes place, i.g. $\omega/\phi \to 3\pi$, $\eta\to3\pi$  \cite{Niecknig:2012sj,Danilkin:2014cra,Guo:2015zqa}.
In presence of many unelastic channels the problem becomes extrimely complicated and has not been solved up to now.
\item {\it Couplings and relative strength}: the addtional qulitative check can be done if the strength of effect is predicted. The calculation of cross-section requires knowledge of the couplings for all vertexes.
\end{enumerate}

\bibliographystyle{h-physrev}
\bibliography{penta}

\end{document}